\documentclass{desyproc}
\begin{document}
\title{Searches for Axion-Like Particles with NGC1275: Current and Future Bounds}

\author{{\slshape Nicholas Jennings}\\[1ex]
Rudolf Peierls Centre for Theoretical Physics, Oxford OX1 3NP}

\contribID{familyname\_firstname}

\confID{13889}  
\desyproc{DESY-PROC-2017-XX}
\acronym{Patras 2017} 
\doi  

\maketitle

\begin{abstract}
Galaxy clusters contain large magnetic fields that make them excellent targets to search for ultralight Axion-Like Particles (ALPs). ALP-photon interconversion imprints quasi-sinusoidal oscillations on the X-ray spectra of point sources in or behind the cluster. The absence of substantial oscillations allows us to place bounds on $g_{a \gamma \gamma}$. Here the bounds from the {\it Chandra} X-ray observations of NGC1275 are presented, as well as those predicted for the {\it Athena} X-ray observatory, due to launch in 2028.
\end{abstract}

\section{ALP-photon conversion in galaxy clusters}

The probability of conversion between Axion-Like Particles (ALPs) and photons in an external magnetic field is a standard result \cite{Sikivie:1983ip, Raffelt:1987im}. ALPs can naturally have very small masses, and for $m_a \lesssim 10^{-12}\,\rm{eV}$ the probability of conversion in a magnetic field takes the form:

\begin{equation}
\qquad P_{a \rightarrow \gamma} = \frac{1}{2}A^2\bigg(\frac{\omega}{\rm{keV}}\bigg)^2\sin^2 \left( C \frac{\rm{keV}}{\omega} \right),
\label{conversion}
\end{equation}
where $A \propto B_{\perp} g_{a \gamma \gamma}/n_e$ for $B_{\perp}$ the magnetic field  perpendicular to the ALP wave vector, $g_{a \gamma \gamma}$ the ALP-photon coupling and $n_e$ the electron density, and $C \propto n_e L$ for domain length $L$. This equation holds for $A \ll 1$, which is generally true in galaxy clusters, where magnetic fields are $\mathcal{O}(\mu\rm{G})$ with coherence lengths $\mathcal{O}(10\,\rm{kpc})$, and electron densities $\mathcal{O}(10^{-3}\,\rm{cm}^{-3})$. 

For these parameters, the ALP-photon conversion probability will imprint quasi-sinusoidal oscillations on the spectrum of a source in the energy range 1-10 keV~\cite{1304.0989, 1312.3947, 1509.06748}. Equation~\ref{conversion} shows that oscillations will be small and rapid at low energies, with increasing amplitude and period at higher energies. The lack of information about the 3D structure of intracluster magnetic fields precludes a precise model of these oscillations; however, their absence in the X-ray spectra of point sources in or behind clusters can constrain $g_{a \gamma \gamma}$~\cite{0902.2320,1305.3603}.

Figure~\ref{Fig:ALPPhotonConversion} illustrates the energy-dependent survival probability for a photon passing across 300 domains of a magnetic field model for the Perseus Cluster, which has an estimated central magnetic field value of $25\,\mu$G \cite{0602622}. The Perseus Cluster hosts a very bright Active Galactic Nucleus (AGN) in its central galaxy NGC1275, whose intrinsic spectrum dominates the background cluster emission and is well described by an absorbed power law \cite{Yamazaki, Fabian:2015kua}, making it an ideal target for ALP searches.

\section{Deriving bounds on $g_{a \gamma \gamma}$}
\label{bounds}

We use a tangled, random domain model for the magnetic field in Perseus, with a central magnetic field value of $B_0 = 25\,\mu $G, following \cite{0602622}. The electron density distribution in Perseus~\cite{Churazov:2003hr}:

\begin{figure}[t]
\centerline{\includegraphics[width=1\textwidth]{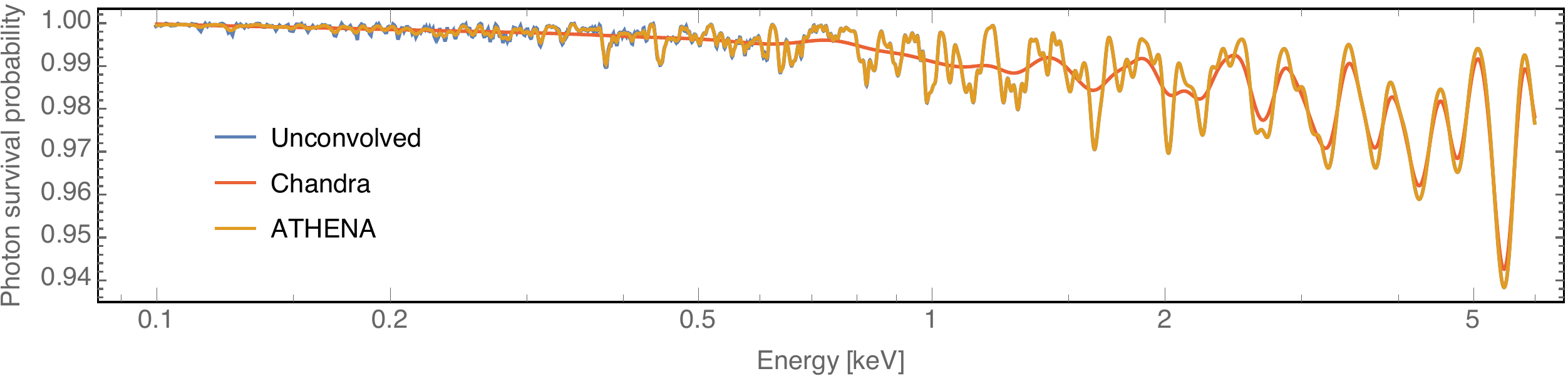}}
\caption{A randomly generated photon survival probability along the line of sight from NGC1275 to us: unconvolved (blue), convolved with a Gaussian with FWHM 150\,eV (the energy resolution of {\it Chandra}'s ACIS-I detector (red)) and 2.5\,eV for {\it Athena}'s X-IFU detector (orange). A central magnetic field of $B_0 = 25\,\mu \rm{G}$ was used, with ${g_{a\gamma\gamma} = 5 \times 10^{-13}\, {\rm GeV}^{-1}}$. At energies $<2$\,keV {\it Chandra} is unable to resolve oscillations, while {\it Athena's} sensitivity is almost indistinguishable from the unconvolved case.}\label{Fig:ALPPhotonConversion}
\end{figure}

\begin{equation}
\qquad n_{e} (r) = \frac{3.9 \times 10^{-2}}{ [ 1 + (\frac{r}{80 \, {\rm kpc}})^2]^{1.8}} +  \frac{4.05 \times 10^{-3}}{ [ 1 + (\frac{r}{280 \, {\rm kpc}})^2]^{0.87}} \, {\rm cm}^{-3},
\end{equation}
determines the radial scaling $B(r) \propto n_{e}(r)^{0.7}$, based on the most conservative exponent value inferred from the cool core cluster A2199~\cite{Vacca:2012up}. We generate the magnetic field over 300 domains, whose lengths are drawn from a Pareto distribution between $3.5~\rm{kpc}$ and $10~\rm{kpc}$ with power 2.8, also motivated by A2199. In each domain the magnetic field and electron density are constant, with a random direction of ${\bf B}$. We generate 50 different configurations for each value of $g_{a \gamma \gamma}$, and calculate the survival probability of a photon passing through (see \cite{1312.3947}).

We multiply this survival probability by an absorbed power law plus thermal background:

\begin{equation}
\qquad F_{0}(E) = (A E^ {- \gamma} + \mathtt{BAPEC}) \times e^{-n_{H} \sigma(E,z)},
\label{AGN}
\end{equation}
where $A$ and $\gamma$ are the amplitude and index of the power law, $E$ is the energy, $n_H$ is the equivalent hydrogen column, $\sigma(E, z)$ is the photo-electric cross-section at redshift $z$, and $\mathtt{BAPEC}$ is the standard plasma thermal emission model. We use parameters derived from the best fit values from the data. For each of these 50 models, we simulate 10 fake PHAs, giving a total of 500 fake data sets. We fit the fake data to the AGN spectrum model \emph{without ALPs} and calculate the resulting reduced $\chi^2$. If $\chi_{fake}^2 < \chi_{data}^2$ (fit to the same no-ALP spectrum model) for fewer than 5\% of the configurations, then that value of $g_{a \gamma \gamma}$ is excluded at 95\% confidence. We repeat the process for different $g_{a \gamma \gamma}$ until this condition no longer holds, producing the bound.

\pagebreak

\section{Bounds from Chandra data of NGC1275}

\begin{wraptable}{r}{0.65\textwidth}
\begin{tabular}{|r|c|c|}
\hline
& {\it Athena} (X-IFU) & {\it Chandra} (ACIS-I)\\
\hline\hline
Energy range & 0.2--12 keV & 0.3--10 keV\\ \hline
Energy resolution & 2.5 eV & 150 eV\\ 
 at 6 keV & & \\ \hline
Spatial resolution & 5 arcsec & 0.5 arcsec\\ \hline
Time resolution & 10~$\mu$s & 0.2 s\\
& & (2.8~ms single row)\\ \hline
Effective area & 2~m$^2$ @ 1 keV & 600~cm$^2$ @ 1.5 keV \\ \hline
\end{tabular}
\caption{Parameters taken from the {\it Athena} Mission Proposal and the {\it Chandra} Proposer's Guide.}
\label{satellites}
\end{wraptable}

The {\it Chandra} X-ray Observatory is well suited to resolving point sources in galaxy clusters, due to its excellent angular resolution of $0.5''$. Archival data contains a set of 4 observations of NGC1275, taken with the ACIS-I instrument in 2009 (\mbox{OBSID} 11713, 12025, 12033 and 12036). These comprise 230\,000 counts in total, and have the least pileup contamination of all the NGC1275 observations. In order to `clean' these observations further we remove the brightest, most contaminated central pixels from the analysis. The resulting spectrum has modulations $\mathcal{O}(10\%)$. From this data we can constrain ${g_{a\gamma\gamma} \lesssim 1.4 \times 10^{-12}\,{\rm GeV}^{-1}}$ at 95\% confidence, for $B_0 = 25\,\mu \rm{G}$~\cite{Berg:2016ese}. This represents a factor of 3 improvement over the bounds from SN1987A in this mass range~\cite{Payez:2014xsa}. Similar bounds have been derived from observations of M87 \cite{Marsh:2017yvc} and 2E3140 \cite{Conlon:2017qcw}. 

\section{Projected bounds from Athena observations of NGC1275}

\begin{wrapfigure}{r}{0.6\textwidth}
\includegraphics[width=0.6\textwidth]{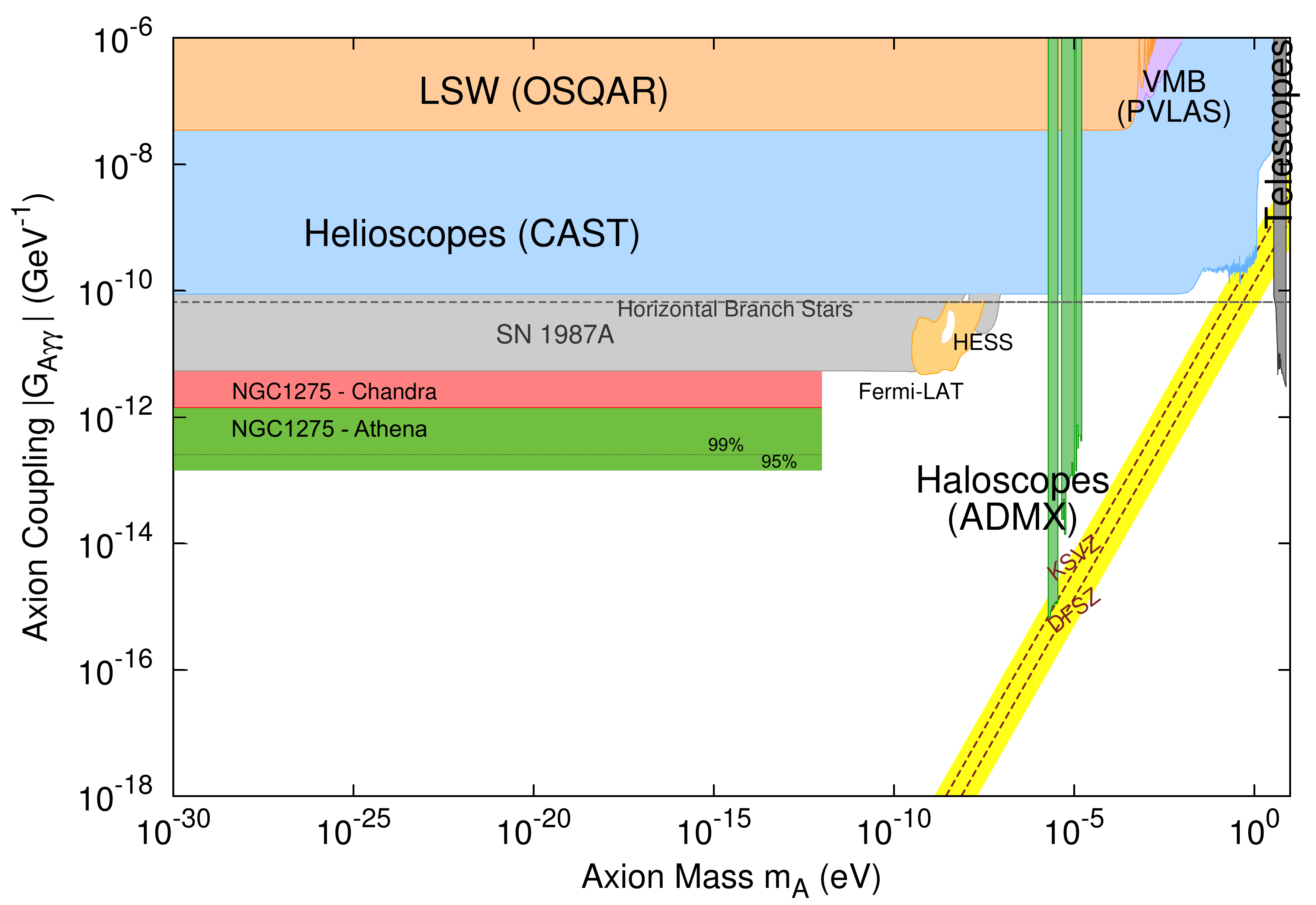}
\caption{Overview of exclusion limits on axion couplings vs mass. Full references can be found in the Particle Data Group review on {\it Axions and other similar particles} \cite{Patrignani:2016xqp}.}\label{fig:ExclusionLimit}
\end{wrapfigure}

The {\it Athena} X-ray observatory, due to launch in 2028, will carry the X-ray Integral Field Unit (X-IFU), a microcalorimeter with an energy resolution of $\sim2.5$~eV \cite{Nandra:2013shg, Gottardi:2016cdx}. This will allow X-IFU to resolve narrow spectral oscillations, while a readout time $\mathcal{O}(10\mu\rm{s})$ will ensure pileup contamination is minimised \cite{Barret:2016ett}, despite a much larger effective area (see Table~\ref{satellites} for a summary of its properties).

In~\cite{Conlon:2017ofb} we simulate the performance of {\it Athena} using the Simulation of X-ray Telescopes ($\mathtt{SIXTE}$) code. It aims to offer an end-to-end simulation, modelling the telescope's vignetting, ARF and PSF, and X-IFU's response, event reconstruction and pileup \cite{2014SPIE.9144E..5XW}. We use $\mathtt{xifupipeline}$ to produce fake data both with and without ALPs. The procedure to determine bounds follows that of Section~\ref{bounds}, the only difference being that -- instead of real data -- we generate 100 fake data sets without ALPs and calculate the average reduced chi-squared. If this is less than the reduced chi-squareds of the data sets with ALPs for fewer than 5\% of cases, then $g_{a \gamma \gamma}$ is excluded at 95\% confidence.

For a simulation of 200\,ks, assuming {\it Athena} does not see any unexpected spectral modulations, we derive a projected bound of ${g_{a\gamma\gamma} \lesssim 1.5 \times 10^{-13}\,\rm{GeV}^{-1}}$ at 95\% confidence, as shown in Figure \ref{fig:ExclusionLimit} alongside the {\it Chandra} bound and published data limits.

\section{Conclusion}

The absence of modulations in the X-ray spectra of point sources in galaxy clusters produce excellent constraints on $g_{a\gamma\gamma}$. The bound ${g_{a\gamma\gamma} \lesssim 1.4 \times 10^{-12}\,\rm{GeV}^{-1}}$ from {\it Chandra} data of NGC1275 is world-leading for $m_a \lesssim 10^{-12}\,\rm{eV}$. {\it Athena}'s groundbreaking energy resolution has the potential to push this bound down to ${g_{a\gamma\gamma} \lesssim 1.5 \times 10^{-13}\,\rm{GeV}^{-1}}$: an order of magnitude improvement, and the strongest cosmology-independent bound in this mass range of any current or currently-planned experiment.

\section*{Acknowledgments}

The work presented here is funded by the European Research Council starting grant `Supersymmetry Breaking in String Theory' (307605). N. Jennings is also funded by STFC.
 

\begin{footnotesize}

\end{footnotesize}


\end{document}